# Studying resonance in the complex charge plane

A. D. Alhaidari

*Physics Department, King Fahd University of Petroleum & Minerals, Box 5047, Dhahran 31261, Saudi Arabia*

email: haidari@mailaps.org

Potential resonances are usually investigated either directly in the complex energy plane or indirectly in the complex angular momentum plane. Another formulation complementing these two is presented in this work. It is an indirect method which studies resonances in a complex charge plane (*Z*-plane). The complex scaling (rotation) method is employed in the development of this formulation. Bound states spectrum and resonance energies are mapped onto a discrete set of poles of a resolvent operator on the real line of the *Z*-plane. These poles move along trajectories as we vary the energy. A finite $L^2$ basis is used in the numerical implementation of the method. Stability of poles and trajectories against variations in all computational parameters is demonstrated. Resonances for a given potential example are calculated and compared with those obtained elsewhere. In this presentation, modest effort will be devoted to mathematical rigor.

## I. INTRODUCTION

Obtaining energy resonances is of prime significance to the study of potential scattering. Several techniques have been developed by many researchers over the years for locating resonances (position and width) for various potential models [1]. The objective of most of these studies is to increase the accuracy of the values obtained and to improve the computational efficiency in locating resonances. All of these investigations are performed in one of two modes; either directly in the complex energy plane or indirectly in the complex angular momentum plane. In the former setting the energy spectrum of the Hamiltonian (poles of the Green's function) consists of three parts: (1) a discrete set of real points on the negative energy axis corresponding to the bound states, (2) the real positive energy line which corresponds to the continuum scattering states, and (3) another discrete set of points in the lower half of the complex energy plane corresponding to resonance states. These are bound-like states that are unstable decaying with a rate that increases with the (negative) value of the imaginary part of the resonance energy. That is, sharp or "shallow" resonances (those that are located below and close to the real energy axis) decay much slower than broad or "deep" resonances which are located below and far away from the real energy axis. Figure 1 shows such a typical structure, which is associated with the potential

$$V(r) = 5e^{-(r-\frac{7}{2})^2/4} - 8e^{-r^2/5} \qquad (1.1)$$

On the other hand, resonances could also be studied in the complex angular momentum plane by locating the Regge poles [2] for a given energy and investigating their trajectories as the (complex) energy is varied. Resonance energies are those that correspond to points on the trajectories where they cross the real angular momentum axis at non-negative integers ($\ell = 0, 1, 2, ...$). Figure 2a is an example showing the trajectories associated with the potential (1.1) for a given range of energies with a fixed negative imaginary part. The Figure shows the lowest five trajectories. It indicates the presence of an $\ell = 3$ resonance. Figure 2b is for the same system, but the energies are real.

The basic principle underlying the various numerical methods for studying resonances, which are implemented within one of these two formulations, is that the position of a resonance is stable against variations in all computational parameters. In this letter the same principle will be used to introduce a third formulation which complements the two that were explained briefly above. The analysis of resonances in this formulation is carried out in a complex charge plane. We employ the complex scaling (rotation) method [3] in the development of this formalism. A simple potential example will be given to illustrate the utility and demonstrate the accuracy of this method. The results obtained will be compared with those found elsewhere in the literature. In addition, new resonances are calculated, some of which are very broad going beyond the range of applicability of some present techniques.

## II. COMPLEX SCALING IN THE CHARGE PLANE

By complex scaling or complex rotation we mean the transformation of the radial coordinate $r$ as

$$r \to re^{i\theta} \tag{2.1}$$

where $\theta$ is a real angular parameter. The Green's function (resolvent operator) is formally defined as $G^\theta \equiv \left(H^\theta - E\right)^{-1}$, where $E$ is the complex energy and $H^\theta$ is the complex-rotated full Hamiltonian of the system. The effect of this transformation on the pole structure of the Green's function in the complex energy plane consists of the following: (1) the discrete bound state spectrum which lies on the negative energy axis remains unchanged; (2) the branch cut (the discontinuity) along the real positive energy axis rotates clockwise by the angle $2\theta$; (3) resonances in the lower half of the complex energy plane located in the sector which is bounded by the new rotated cut line and the positive energy axis get exposed and become isolated. However, due to the finite size of the basis used in performing the calculation, the matrix representation of the Hamiltonian is finite resulting in a discrete spectrum. Consequently, the rotated cut line gets replaced by a string of interleaved poles and zeros of the finite Green's function which tries to mimic the cut structure. This method will now be developed to make it suitable for implementation in the complex charge plane.

In the atomic units $\hbar = m = 1$, the one-particle wave equation for a spherically symmetric potential $V(r)$ in the presence of the Coulomb field reads as follows

$$\left(H - E\right)\chi = \left[-\frac{1}{2}\frac{d^2}{dr^2} + \frac{\ell(\ell+1)}{2r^2} + \frac{Z}{r} + V(r) - E\right]\chi = 0, \tag{2.2}$$

where $\ell$ is the orbital angular momentum and $Z$ is the electric charge in units of $e$. $\chi(r)$ is the wavefunction which is parameterized by $\ell$, $Z$, $E$ and the potential parameters. Equation (2.2) could be rewritten as $\left(\hat{H} - Z\right)\hat{\chi} = 0$, where

$$\hat{H} = \frac{r}{2}\frac{d^2}{dr^2} - \frac{\ell(\ell+1)}{2r} + rE - rV(r) \equiv \hat{H}_0 + \hat{V}, \tag{2.3}$$

and $\hat{V} \equiv -rV(r)$. The continuum could be discretized by taking $\hat{\chi}$ as an element in an $L^2$ space with a complete basis set $\{\phi_n\}$. The integration measure in this space is $dr/r$. We parameterize the basis by a scale parameter $\lambda$ as $\{\phi_n(\lambda r)\}$. The following realization of the basis functions is compatible with the domain of the operator $\hat{H}$ and satisfies the boundary conditions (at $r = 0$ and $r \to \infty$):

$$\phi_n(\lambda r) = A_n x^\alpha e^{-x/2} L_n^\nu(x), \qquad (2.4)$$

where $x = \lambda r$, $\alpha > 0$, $\nu > -1$, and $n = 0,1,2,....$ $L_n^\nu(x)$ is the Laguerre polynomial of order $n$ and $A_n$ is the normalization constant $\sqrt{\Gamma(n+1)/\Gamma(n+\nu+1)}$. The choice $2\alpha = \nu + 1$ makes the basis set $\{\phi_n\}$ orthonormal. The matrix representation of the "reference" operator $\hat{H}_0$ in this basis is written as

$$\left(\hat{H}_0\right)_{nm} = \langle \phi_n(x) | \frac{\lambda}{2} x \frac{d^2}{dx^2} - \ell(\ell+1)\frac{\lambda}{2x} + \frac{E}{\lambda} x | \phi_m(x) \rangle \qquad (2.5)$$

Consequently, the action of the transformation (2.1) on $\hat{H}_0$ is equivalent to

$$\lambda \to \lambda e^{-i\theta} \qquad (2.6)$$

In the manipulation of (2.5) we use the differential equation, differential formulas and three-term recurrence relation of the Laguerre polynomials [4]. As a result, we obtain the following elements of the matrix representation of the reference operator

$$\left(\hat{H}_0\right)_{nm} = \lambda\left(\frac{E}{\lambda^2} - \frac{1}{8}\right)(2n+\nu+1)\delta_{n,m} - \lambda\left(\frac{E}{\lambda^2} + \frac{1}{8}\right)\sqrt{n(n+\nu)}\delta_{n,m+1}$$
$$-\lambda\left(\frac{E}{\lambda^2} + \frac{1}{8}\right)\sqrt{(n+1)(n+\nu+1)}\delta_{n,m-1} + \frac{\lambda}{8}\left[\nu^2 - (2\ell+1)^2\right]\left(x^{-1}\right)_{nm} \qquad (2.7)$$

where the symmetric matrix $\left(x^{-1}\right)_{nm} = \frac{1}{2\nu}\left(\frac{A_n}{A_m} + \frac{A_m}{A_n}\right)$. To simplify this representation we take $\nu = 2\ell + 1$ resulting in a tridiagonal matrix representation for $\hat{H}_0$. Now, the only remaining quantity that is needed to carry out the calculation is the matrix elements of the "potential" term $\hat{V}$ defined in Eq. (2.3). This is obtained by evaluating the integral

$$\hat{V}_{nm} = \int_0^\infty \phi_n(\lambda r)\left[-rV(r)\right]\phi_m(\lambda r)\frac{dr}{r}$$
$$= \frac{-1}{\lambda} A_n A_m \int_0^\infty x^\nu e^{-x} L_n^\nu(x) L_m^\nu(x) \left[xV(x/\lambda)\right] dx \qquad (2.8)$$

The evaluation of such integral is almost always done numerically. We use the Gauss quadrature approximation [5] which gives

$$\hat{V}_{nm} \cong \frac{-1}{\lambda} \sum_{k=0}^{N-1} \Lambda_{nk} \Lambda_{mk} \left[\mu_k V(\mu_k/\lambda)\right], \qquad (2.9)$$

for some large enough integer $N$. $\mu_k$ and $\{\Lambda_{nk}\}_{n=0}^{N-1}$ are the $N$ eigenvalues and respective eigenvectors of the $N \times N$ tridiagonal symmetric $J$ matrix whose elements are

$$J_{n,n} = 2n+\nu+1, \quad J_{n,n+1} = -\sqrt{(n+1)(n+\nu+1)}. \qquad (2.10)$$

In the following section, these findings will be used in locating and analyzing the resonance structure for the potential $V(r) + Z/r$ in the complex $Z$-plane. An example will be given to illustrate the utility and accuracy of the proposed method.

### III. STUDYING RESONANCE IN THE COMPLEX Z-PLANE

The system described by Eq. (2.2) could be studied by investigating an equivalent one described by the equation $(\hat{H} - Z)\hat{\chi} = 0$. However, this equivalence is only an approximation that improves with the increase in the size of the basis, *N*. Our investigation of the latter system is made by using the method of complex rotation which is implemented, as explained above, by applying the transformation $\lambda \to \lambda e^{-i\theta}$ on the matrix representations (2.7) and (2.9). The resulting complex eigenvalues $\{Z_n\}_{n=0}^{N-1}$ of $\hat{H}^\theta$ are the poles of the finite Green's function $\hat{G}^\theta = \left(\hat{H}_0^\theta + \hat{V}^\theta - Z\right)^{-1}$. The subset of these poles that are stable (in the complex *Z*-plane) against variations in the parameters $\lambda$ and $\theta$ are the ones that are physically significant. However, $\theta$ must always be larger than a minimum angle needed to expose the poles of interest. The branch cut of the Green's function $\hat{G}(Z)$ is located on the negative *Z*-axis. Complex scaling rotates this cut clockwise through the angle $\theta$ exposing the relevant poles. This behavior could be understood by comparing it with the corresponding one in the complex energy plane and noting that (i) the relative sign of *Z* to that of *E* in the Hamiltonian (2.2) is negative, and that (ii) the length dimensions of *E* and *Z* are the same as that of $\lambda^2$ and $\lambda$, respectively. As we vary the energy (generally, complex) the poles of $\hat{G}^\theta$ move along trajectories in the complex *Z*-plane. The points where the *stable* trajectories cross the real *Z*-axis correspond to resonances. For elementary charged particles, the relevant crossings are those at $Z = 0, \pm 1, \pm 2, \ldots$.

The union of all of the sets of eigenvalues $\{Z_n(E)\}_{n=0}^{N-1}$ for a given range of (complex) values of *E* produces *N* trajectories in the *Z*-plane. The physically relevant ones are those which are stable against variations (around a plateau) in all computational parameters. It should, however, be noted that the ordering of these eigenvalues by the index *n* is computationally-dependent and is not necessarily the same for two different values of *E*. Consequently, one has to plot all of the eigenvalues for each *E* in the energy range to see the full picture. To illustrate these findings we consider, as an example, the potential function

$$V(r) = 7.5 r^2 e^{-r}, \tag{3.1}$$

which has been studied frequently in the literature [6-10]. Figure 3a shows the lowest *s*-wave trajectories for real energies. These trajectories do not cross the real *Z*-axis but only coincide with it for real negative energies. The same graph is repeated in Fig. 3b, but now the energy has a non-vanishing imaginary part. It shows several crossings at, or near, $Z = -8, -4, 9$ indicating resonance. In practice, we vary the imaginary part of *E* until it produces trajectory plots that have at least one of its branches crossing the real *Z*-axis at an integral value. Subsequently, one zooms in at the crossings to refine the search. A numerical algorithm using, for example, bisection or Newton-Raphson routines [11] could be developed to automate the search.

In our calculation, we start with a rough estimate of resonance values obtained by the complex rotation method [3] for the potential (3.1). The extent of accuracy of the proposed formalism is then demonstrated by improving on these values for a given moderate size, *N*, of the basis. Table I compares values found in the literature to those obtained in this work. Other resonances are also obtained in Table II some of which are very broad. Stability of these results against variations in $\lambda$ and $\theta$ (for a basis size *N* = 200) goes up to the tenth decimal place.

Nonetheless, the merit of the present approach is not in the accuracy of the results obtained but rather in the global picture it gives for the overall behavior of resonance and its structure. For this it is endowed with formal and computational analogies to the Regge poles and Regge trajectories in the complex $\ell$-plane.

## ACKNOWLEDGMENT

The author is grateful to H. A. Yamani and M. S. Abdelmonem for fruitful discussions that gave motivation to this work.

**TABLES CAPTIONS:**

**Table I**: Resonance energies ($E = \mathcal{E}_r - i\Gamma/2$) for the potential $7.5r^2e^{-r} + Z/r$. Our results are compared with those found in the cited references. Stability of our calculation is against a substantial range of variations in $\lambda$ (~ 5 to 100 a.u.) and $\theta$ (~ 0.5 to 1.0 radians). The accuracy is with respect to a basis dimension $N = 200$.

**Table II**: Another list of resonances for the same potential of Table I and for several values of $Z$ and $\ell$. These were obtained by searching the complex $Z$-plane with a basis size $N = 200$.

**FIGURES CAPTIONS:**

**Fig. 1**: The poles (dots) and discontinuity (line) of the *p*-wave Green's function for the system whose potential is given by Eq. (1.1). One bound state and eleven resonances (two of them are sharp) are shown.

**Fig. 2a**: The trajectory of Regge poles in the complex $\ell$-plane for the system whose potential is given by Eq. (1.1). Along these trajectories the real part of the energy varies smoothly from 3.0 to 10.0 a.u. while the imaginary part is fixed at −2.0 a.u. The graph indicates the presence of a resonance where the trajectory crosses the real line at $\ell = 3$. In fact, our calculation gives the following value for this resonance: $E = 7.09172304 - i\, 2.00173429$ a.u.

**Fig. 2b**: Regge trajectories plot for the same system of Fig. 2a except that the imaginary part of the energy vanishes. Consequently, no crossing of the real line anywhere but only coincidence with it is allowed for real negative energies.

**Fig. 3a**: The trajectory of the poles of the *s*-wave Green's function $\hat{G}^\theta(Z)$ in the complex $Z$-plane for the system whose potential is $7.5r^2e^{-r} + Z/r$. Along these trajectories the energy is real and varies smoothly from 0.0 to 10.0 a.u. These trajectories do not cross the real $Z$-axis but only coincide with it for real negative energies.

**Fig. 3b**: Same as Fig. 3a except that the energy is now complex with a negative imaginary part which is kept constant at −3.0 a.u. The crossings at, or near, $Z = -8, -4, 9$ signifies resonances which are calculated to be: $1.287274955 - i\, 2.971759279$, $3.125581370 - i\, 3.023378045$, and $9.733679948 - i\, 2.988524088$ a.u., respectively.

**Table I**

| $Z$ | $\mathcal{E}_r$ (a.u.) | $\Gamma$ (a.u.) | Reference |
|---|---|---|---|
| 0 | 3.42639 | 0.025549 | [6] |
|   | 3.4257 | 0.0256 | [7] |
|   | 3.426 | 0.0256 | [8] |
|   | 3.426 | 0.0258 | [9] |
|   | 3.426390331 | 0.025548962 | [10] |
|   | 3.426390310 | 0.025548961 | this work |
| 0 | 4.834806841 | 2.235753338 | [10] |
|   | 4.834806841 | 2.235753338 | this work |
| 0 | 5.277279780 | 6.778106356 | [10] |
|   | 5.277279864 | 6.778106591 | this work |
| −1 | 1.7805 | $9.58 \times 10^{-5}$ | [9] |
|   | 1.780524536 | $9.5719 \times 10^{-5}$ | [10] |
|   | 1.780524536 | $9.57194 \times 10^{-5}$ | this work |
| −1 | 4.101494946 | 1.157254428 | [10] |
|   | 4.101494946 | 1.157254428 | this work |
| −1 | 4.663461099 | 5.366401539 | [10] |
|   | 4.663461097 | 5.366401540 | this work |

**Table II**

| | $\ell = 0$ | | $\ell = 1$ | | $\ell = 2$ | |
|---|---|---|---|---|---|---|
| Z | $\mathcal{E}_r$ (a.u.) | $\Gamma$ (a.u.) | $\mathcal{E}_r$ (a.u.) | $\Gamma$ (a.u.) | $\mathcal{E}_r$ (a.u.) | $\Gamma$ (a.u.) |
| 0 | 5.064929608 | 11.952069576 | 5.4277422973 | 9.280688974 | 5.7936930648 | 6.660951732 |
| | 4.268860299 | 17.433816868 | 5.3604696511 | 4.394482330 | 5.5029439380 | 11.843122555 |
| | 2.9477816003 | 23.061029462 | 4.8877690564 | 14.623615321 | 5.4913453119 | 2.1072915737 |
| | 1.147183738 | 28.738014274 | 4.6466344207 | 0.6505890286 | 4.6527742298 | 17.324910417 |
| | −1.096688979 | 34.402043587 | 3.8017984630 | 20.187028641 | 3.2892316413 | 22.949471910 |
| | −3.75414412 | 40.00901499 | 2.2189063561 | 25.846441901 | 1.4529914754 | 28.624988427 |
| | −6.800304 | 45.52631 | 0.1785773686 | 31.524321640 | −0.821534539 | 34.290493164 |
| | −10.21 | 50.93 | −2.28696411 | 37.167523861 | −3.50540256 | 39.9020901 |
| | | | −5.150992 | 42.73742 | −6.5742 | 45.4265 |
| | | | −8.39 | 48.20 | −10.0 | 51 |
| −1 | 4.561151055 | 10.413396043 | 4.932942955 | 8.233838942 | 5.3006134903 | 5.884714861 |
| | 3.849197760 | 15.816567000 | 4.750053489 | 3.505579849 | 5.0886414686 | 10.943326210 |
| | 2.5933485088 | 21.387481635 | 4.476218206 | 13.477261705 | 4.9005161468 | 1.5675070165 |
| | 0.844647138 | 27.021909107 | 3.8480016348 | 0.2753844592 | 4.3001798532 | 16.341585409 |
| | −1.356846588 | 32.653003793 | 3.4539968253 | 18.968736871 | 2.986244924 | 21.903035992 |
| | −3.978854424 | 38.23413033 | 1.9216035248 | 24.572340524 | 1.1907356869 | 27.527998140 |
| | −6.9947 | 43.73092 | −0.0777332579 | 30.205316536 | −1.0497907711 | 33.151641912 |
| | −10.38 | 49.1 | −2.509315965 | 35.81169783 | −3.70488468 | 38.727911891 |
| | | | −5.34469 | 41.35108 | −6.749011 | 44.22222 |
| | | | −8.56 | 46.8 | −10.16 | 49.61 |
| +1 | 5.867437031 | 8.158768524 | 5.9409036759 | 5.2802624063 | 6.2697298719 | 7.448168504 |
| | 5.569681242 | 3.422105119 | 5.902424814 | 10.320417212 | 6.0571881775 | 2.6905607459 |
| | 5.545052842 | 13.452021982 | 5.3718654458 | 1.1462418923 | 5.905180463 | 12.745647181 |
| | 4.666831409 | 19.017323601 | 5.2840834258 | 15.761247384 | 4.9960630849 | 18.307602572 |
| | 4.594490160 | 0.2578906726 | 4.1372090078 | 21.396522076 | 3.584578573 | 23.993734433 |
| | 3.2827571775 | 24.707275525 | 2.5056697545 | 27.112401950 | 1.7086936696 | 29.719160385 |
| | 1.4325150596 | 30.432561004 | 0.4256981801 | 32.836153732 | −0.599061162 | 35.426371269 |
| | −0.8515999765 | 36.134066609 | −2.0727213505 | 38.51720191 | −3.31110673 | 41.07340578 |
| | −3.542588981 | 41.77027553 | −4.96449 | 44.11855 | −6.404053 | 46.62823 |
| | −6.61738 | 47.31057 | −8.2283 | 49.6 | −9.86 | 52.07 |
| | −10.056 | 52.73 | | | | |

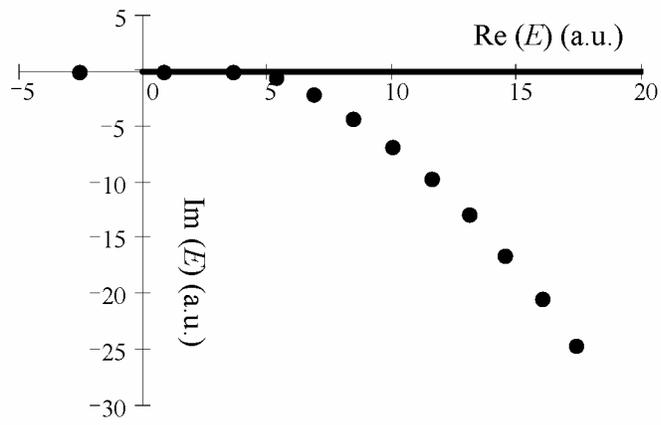

Fig. 1

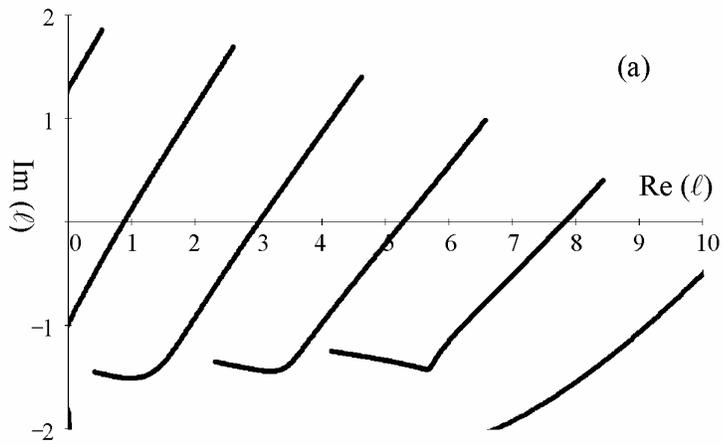

Fig. 2a

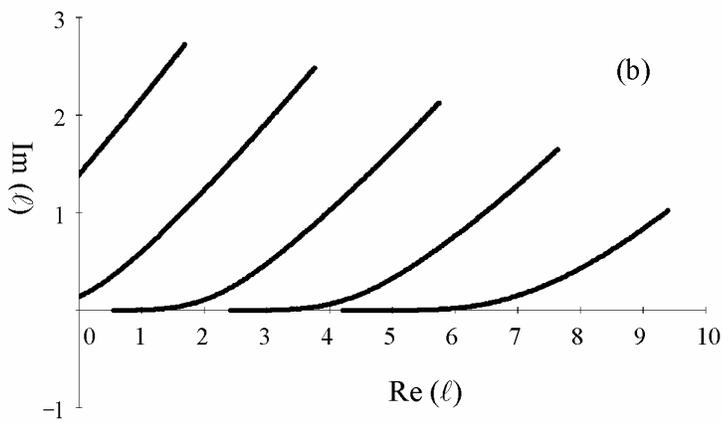

Fig. 2b

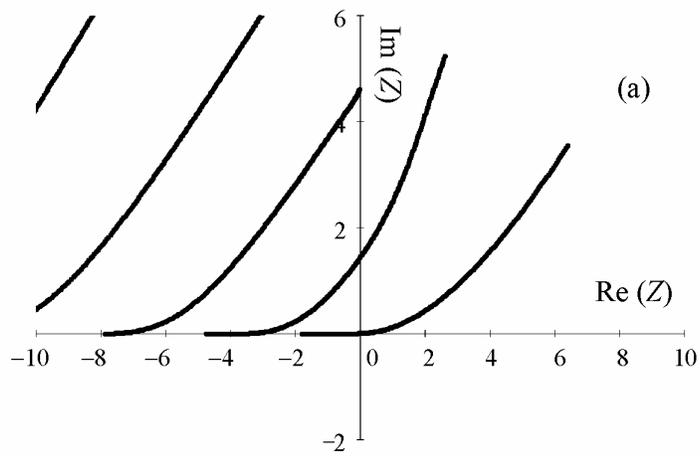

Fig. 3a

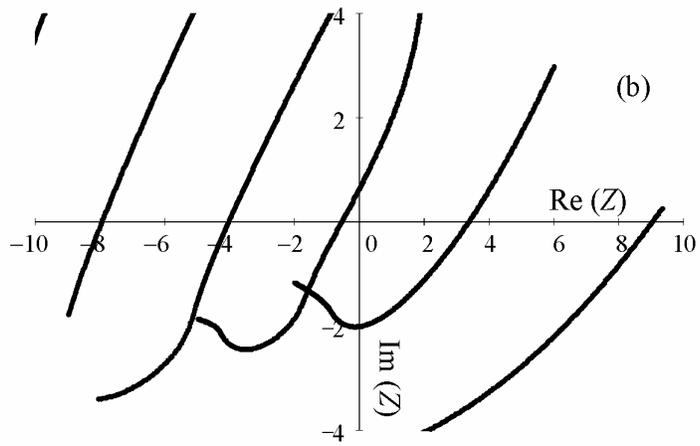

Fig. 3b